\documentclass[english]{article}
\usepackage[T1]{fontenc}
\usepackage[latin9]{inputenc}
\usepackage{geometry}
\usepackage{color}
\usepackage{verbatim}
\usepackage{amsmath}
\usepackage{amssymb}
\usepackage{graphicx}
\usepackage{setspace}
\usepackage{authblk}
\usepackage{multirow}
\usepackage{siunitx}
\makeatletter

\date{}
\@ifundefined{showcaptionsetup}{}{%
\PassOptionsToPackage{caption=false}{subfig}}
\usepackage{subfig}
\makeatother
\usepackage{babel}
\usepackage{float}
\usepackage{multirow}
\usepackage{enumitem}
\usepackage{longtable}
\usepackage{textcomp}

\begin{document}
\title{\textcolor{black}{Crystallographic design of intercalation materials}}
\author{Ananya Renuka Balakrishna\thanks{Corresponding author: renukaba@usc.edu}}
\affil{\small{Aerospace and Mechanical Engineering, University of Southern California, Los Angeles, CA 90089, USA}}

\maketitle
\begin{abstract}
Intercalation materials are promising candidates for reversible energy storage and are, for example, used as lithium-battery electrodes, hydrogen-storage compounds, and electrochromic materials. An important issue preventing the more widespread use of these materials is that they undergo structural transformations (of up to $\sim10\%$ lattice strains) during intercalation, which expand the material, nucleate microcracks, and, ultimately, lead to material failure. Besides the structural transformation of lattices, the crystallographic texture of the intercalation material plays a key role in governing ion-transport properties, generating phase separation microstructures, and elastically interacting with crystal defects. In this review, I provide an overview of how the structural transformation of lattices, phase transformation microstructures, and crystallographic defects affect the chemo-mechanical properties of intercalation materials. In each section, I identify the key challenges and opportunities to crystallographically design intercalation compounds to improve their properties and lifespans. I predominantly cite examples from the literature of intercalation cathodes used in rechargeable batteries, however, the identified challenges and opportunities are transferable to a broader range of intercalation compounds. 
\end{abstract}

\section*{Introduction}\label{section: Introduction}

Intercalation is the reversible insertion of active ions (or molecules) into a materials' lattice structure. This reversible insertion is often accompanied by a simultaneous change to the host materials' lattice structure and/or its physical properties (i.e., optical, thermal, electronic) \cite{sood2021electrochemical}. This intrinsic coupling between ionic (or molecular) insertion and physical properties makes intercalation materials well-suited for next-generation applications, such as electrodes in lithium batteries \cite{padhi1997phospho}, electrochromic materials in smart windows \cite{chen2019gate, liu2020spontaneous}, and ion-exchange membranes in water desalination devices \cite{pothanamkandathil2020electrochemical}. Despite the technological relevance of intercalation materials, their widespread use is plagued by chemo-mechanical challenges that limit material performance and lifespans \cite{lewis2019chemo, van2013understanding, radin2017narrowing}. 

The chemo-mechanical challenges in intercalation materials are closely associated with their crystallographic texture \cite{van2013understanding, rong2015materials}. Inserting active ions into an intercalation material typically induces an abrupt change to its lattice parameters. This structural transformation of lattices---a characteristic feature of first-order phase transformation materials---induces macroscopic volume changes and generates lattice misfit strains in the material \cite{koerver2017capacity, xiang2017accommodating, radin2017narrowing, bucci2018mechanical}. With repeated intercalation, these volume changes and internal stresses weaken the structural integrity of the material and shorten its lifespan. In addition to structural transformation of individual lattices, the crystallographic texture of the intercalation material plays a dominant role in governing ion-transport properties \cite{van2013understanding}.  For example, intercalating ions in LiCoO$_2$ (a commonly used cathode) preferentially diffuse parallel to the (001) crystallographic planes. This preferential diffusion constrains ion migration to specific crystallographic channels and thus limits rapid charge/discharge properties \cite{saber2021role, balke2010nanoscale}. The presence of crystal defects, such as grain boundaries and edge dislocations, further affects ion diffusion pathways \cite{xu2020charge} and contributes to the irreversible cycling of intercalation compounds \cite{ulvestad2015topological}. This interplay between the crystallographic texture and material properties leads us to a central question: Can we design the crystallographic texture of intercalation materials to improve their performance and lifespan?

Take shape memory alloys---another family of phase transformation materials---which share commonalities with structural transformations in intercalation materials. For example, the Ti-Ni-Cu shape memory alloy is accompanied by $\sim 8\%$ lattice strains during phase transformation, which is comparable to lattice strains in common intercalation materials. This large transformation strain contributes to material fatigue in the shape memory alloy after a few stress strain cycles. In a recent work, researchers crystallographically engineered this alloy's composition (through systematic doping) to precisely satisfy a specific lattice geometry \cite{chluba2015ultralow, song2013enhanced}. By doing so, they dramatically enhanced the reversible cycling of the shape memory alloy: the resulting alloy not only showed negligible fatigue after ten million cycles but did so despite its large transformation strain and despite being subjected to a load of over 300MPa. Besides shape memory alloys, researchers have crystallographically engineered highly reversible microstructures in other phase transformation materials, including ferroelectrics \cite{wegner2020correlation}, fracture-resistant ceramics \cite{pang2019reduced}, and semiconductors \cite{liang2020tuning}. These advances in phase transformation materials motivate a new line of research in which the existing intercalation materials can be similarly designed to achieve dramatic enhancement in material lifespan.

In another line of research, the crystallographic texture of intercalation materials can be locally engineered, such as by seeding vacancies or doping grain boundaries, to improve material properties \cite{yan2020perspective}. For example, clusters of vacancy defects have been shown to mediate ion diffusion in intercalation materials \cite{van2010vacancy, van2013understanding, islam2014lithium}. Similarly, small changes to the local structure and chemistry of a grain boundary (referred to as complexions \cite{cantwell2014grain}) can significantly alter battery materials' mechanical and transport properties. Other defects, such as dislocations and anti-site defects, which are often viewed as deleterious features, can be tailored to enhance reversible cycling \cite{hong2019mechanism}. These examples demonstrate potential opportunities to significantly improve material properties by locally designing the crystallographic texture of intercalation materials.

My central aim is to provide an overview of how crystal symmetry and crystallographic microstructures affect intercalation material properties, and how we can intervene and design these crystallographic features to significantly enhance material behavior. I address this central theme in three parts: First, I introduce the types of structural transformations of unit cells during intercalation and highlight potential opportunities to design facile structural changes. Second, I review the different phase transformation microstructures that form in intercalation materials and share my perspective on crystallographically designing these microstructural patterns. Finally, I identify various forms of crystal defects and imperfections commonly observed in crystalline solids and discuss the challenges in tailoring these defects to enhance material performance. In this review, I predominantly cite examples from the literature on intercalation cathodes, however, the challenges and potential opportunities I identify are transferable to a broader range of intercalation compounds.

\section*{Structural Transformation}\label{section: Structural transformation}

\begin{figure}
    \centering
    \includegraphics[width=0.6\textwidth]{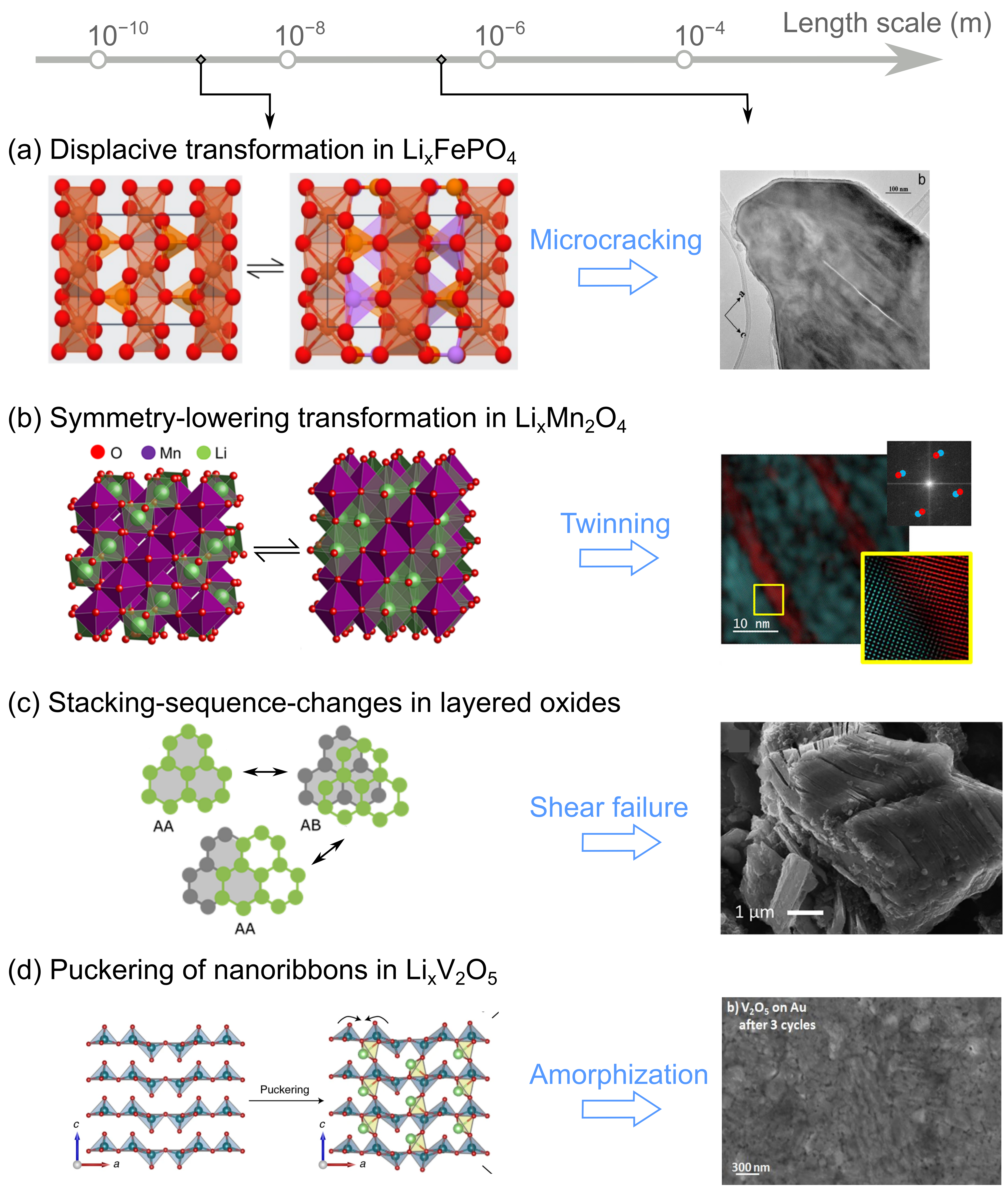}
    \caption{Structural transformations of individual lattices generate at the atomic scale ($\sim10^{-9}$m) generate significant lattice misfit at the continuum scale ($\sim100$nm). These structural transformations manifest in various forms and are directly linked to structural degradation of intercalation materials as follows. (a) The displacive transformation of individual lattices in Li$_x$FePO$_4$ induces coherency stresses at the FePO$_4$/Li$_x$FePO$_4$ phase boundary, which in turn nucleates microcracks. Reprinted with permission from \cite{chen2006electron, jain2013commentary}. (b) The cubic-to-tetragonal symmetry lowering transformation in Li$_x$Mn$_2$O$_4$ generates a finely twinned mixture that minimizes coherency stresses at the phase boundary. Reprinted with permission from the American Chemical Society \cite{erichsen2020tracking}. (c) The stacking-sequence-change transformation of lattices in layered oxides contributes to lattice-invariant shear resulting in microcracking of the intercalation compound. Reprinted with permission from \cite{alvarado2017improvement, radin2017role} and the American Chemical Society. (d) The polyhedral chains (nanoribbons) in Li$_x$V$_2$O$_5$ unfold and pucker during intercalation resulting in a disordered (or amorphous) phase. Reprinted from \cite{zhang2021film} with permission from Elsevier. }
    \label{fig: Structural transformation}
\end{figure}

The reversible intercalation of active species into a host-electrode often induces an abrupt structural transformation of its crystal lattices.\footnote{These abrupt changes to the lattice geometry accompanying a first-order phase transformation differ from a gradually changing lattice geometry accompanying a stoichiometric chemical expansion of materials (e.g., Fluorite-Ceria) \cite{bishop2014chemical}.} These structural changes range from displacive transformations (in which atoms undergo cooperative movement resulting in changes to lattice parameters \cite{bishop2014chemical}) to transformations accompanied by polyhedral rotations that not only generate significant lattice strains but also affect the electronic structure of the host material \cite{lewis2019chemo, bucci2018mechanical}. These structural transformations impede ion migration, nucleate microcracks, and contribute to the irreversible decay of material properties, see Fig.~\ref{fig: Structural transformation}(a-d) \cite{chen2006electron, kim2019correlating, koerver2017capacity, tealdi2016feeling}. These challenges highlight the need to understand different types of structural transformations, and to investigate whether these transformations can be \textit{crystallographically designed} to enhance material properties. In this section, I review the different types of structural transformations primarily using intercalation cathodes as representative examples and identify key challenges and opportunities to design facile transformations.

\subsection*{Types of Structural Transformation}
A common mode of structural transformation in cathodes is the volumetric dilation of individual lattices on intercalation. For example, inserting lithium into FePO$_4$ (an olivine compound) increases its $a$ and $b$ lattice parameters (while simultaneously decreasing the $c$ parameter) resulting in a net volume increase of $\sim6\%$. This deformation is reversible on lithium extraction, however, can induce microcracking on repeated cycling, see Fig.~\ref{fig: Structural transformation}(a). Both the intercalated and reference lattices (e.g., LiFePO$_4$ and FePO$_4$) belong to the same space group and show no reduction in crystal symmetry.\footnote{Please note that symmetry-preserving structural transformations could still have different lattice geometries.} Similar symmetry-preserving transformations are observed in commonly used electrodes such as layered LiCoO$_2$ and Li-NMC \cite{whittingham2004lithium}, and in lithium battery anodes such as  Li$_4$Ti$_5$O$_{12}$ \cite{verde2016elucidating}.

By contrast, cathode compounds such as the manganese spinel, undergo cooperative Jahn-Teller distortion on intercalation and are characterized by an abrupt and spontaneous symmetry breaking of the unit cell \cite{rossouw1990structural, erichsen2020tracking}. For example, electrochemical insertion of lithium into a cubic LiMn$_2$O$_4$ transforms it to a tetragonal Li$_2$Mn$_2$O$_4$ unit cell. This cubic-to-tetragonal transformation is often not entirely reversible thus limiting the uptake of active ions \cite{huang1999correlating, thackeray2018quest}. However, this symmetry-lowering transformation is necessary for the material to form finely-twinned microstructures, see Fig.~\ref{fig: Structural transformation}(b). Other transition-metal oxides, such as PNb$_9$O$_{25}$, are also accompanied by a tetragonal distortion (from Jahn-Teller effect) of its lattices along the $c-$axis. These structural distortions coupled with the electronic structure of PNb$_9$O$_{25}$ affects the ordered filling of multiple Li ions \cite{saber2021role}. Overall, these abrupt crystallographic changes can induce significant misfit strains between neighboring lattices and collectively generate volume changes. On repeated intercalation, these structural changes manifest as interfacial stresses leading to microcracking at phase boundaries (see Fig.~\ref{fig: Structural transformation}(a)) and delamination at the electrode/electrolyte interfaces \cite{lewis2019chemo}.

Another mode of structural transformation, which is unique to layered electrodes, is the stacking-sequence-change mechanism \cite{radin2017role}. In this type of transformation, intercalation of an active species induces a relative displacement between individual layers of the electrode, however, each layer is topotactically preserved. Fig.~\ref{fig: Structural transformation}(c) shows a schematic illustration of this stacking-sequence-change mechanism: At first, two layers initially positioned at `A' (i.e., AA layer) transform to an AB configuration on intercalation (i.e., atoms in the lower layer shift to position `B'). Next, on extraction of the intercalant species, the AB configuration transforms to the AA configuration; however, this configuration can be different from its initial AA configuration. That is, the reverse transformation (AB $\to$ AA) can result in a relative shift of the two layers. Although these stacking-sequence-change transformations are kinetically unfavorable and rare, they can induce lattice-invariant shearing of electrode particles, leading to characteristic step-formation and microcracking in layered oxides (e.g., Li$_x$CoO$_2$, Li$_x$NiO$_2$, Li$_x$Ni$_y$Mn$_z$Co$_{(1-y-z)}$O$_2$) \cite{amatucci1996coo2, van1998first, croguennec2001structural} and sulfides (e.g., Na$_x$TiS$_2$) \cite{vinckeviciute2016stacking}, see Fig.~\ref{fig: Structural transformation}(c). These microstructural changes in layered compounds impact battery performance by contributing to capacity loss and increased polarization with electrochemical cycling \cite{radin2017role}.

The layered structure of V$_2$O$_5$, another widely studied intercalation compound, undergoes an unfolding type of structural transformation on deintercalation, see Fig.~\ref{fig: Structural transformation}(d) \cite{cocciantelli1991preparation, cocciantelli1992electrochemical}. For example, the $\gamma-$LiV$_2$O$_5$ structure consists of VO$_5$ pyramids that share edges and are connected to form ribbons. These ribbons are then connected by pyramid corners that results in a puckered-type of layers with lithium atoms located between the sheets, see Fig.~\ref{fig: Structural transformation}(d).\footnote{Further intercalation of lithium leads to the formation of a metastable phase that has been examined as a potential cathode candidate for multivalent ion intercalation \cite{andrews2018reversible}.} On deintercalation, the interlayer separation in $\gamma$-LiV$_2$O$_5$ decreases while the interribbon distance increases. This structural transformation is accompanied by an unfolding of the V$_2$O$_5$ layers. On deep discharge of Li$_x$V$_2$O$_5$ (i.e., inserting $x > 2$ Li ions), the cathode becomes increasingly disordered (or amorphous) and contributes to the structural degradation of the material \cite{christensen2018structural, zhang2021film}.

More recently, lithiation induced polyhedral rotations in perovskite cathodes have been reported as another mode of structural transformation \cite{bashian2018correlated}. Perovskite compounds such as ReO$_3$ have an open framework and can accommodate up to two Li ions. However, recent studies have shown that electrochemical insertion induces a correlated rotation of its polyhedral subunits that not only generate significant lattice strains but also affect the electronic structure of the cathode material. These rotational type of distortions severely impede the reversible capacity of cathodes over extended cycling, and demonstrate the need to crystallographically design structural transformations.

These various forms of structural transformations, although they occur at the atomic scale, significantly affect the material's structural integrity at the continuum scale. For example, intercalation cathodes undergoing crystal-symmetry-lowering transformations collectively generate macroscopic volume changes and internal stresses which, respectively, lead to electrode delamination and fracture. Similarly, the stacking-sequence-change in layered cathodes generates a lattice-invariant shear resulting in microcracking of electrodes. The unfolding transformation of polyhedral ribbons in V$_2$O$_5$ and the polyhedral rotations in perovskite compounds also lead to irreversible structural degradation of intercalation compounds. These structural degradations expose fresh active areas of the electrode to the electrolyte and act as a site for side chemical reactions. These parasitic reactions, in turn, over time contribute to the decay in the electrochemical performance of the intercalation material.

\subsection*{Designing structural transformations}

Researchers have developed different strategies to minimize the deleterious effects of structural transformations. In one strategy, called `staging', researchers diffuse the guest species into select crystallographic channels of the intercalation material (e.g., in graphite systems \cite{dahn1982elastic, xu2017recent}), occupying only a fraction of the intercalation sites and thereby reducing internal stresses in the materials. In other strategies, researchers suppress structural transformations by engineering host materials as nanoparticles, by designing mechanical constraints as epitaxial strains\cite{zhang2021film} or coherency stressses \cite{cogswell2012coherency}, and by operating intercalation materials within specific voltage windows. While these strategies delay chemo-mechanical degradation of intercalation materials and have improved their lifespans, they do so at the cost of the material's energy storage capacity and performance \cite{radin2017narrowing}.

Recent advances in experimental synthesis techniques have opened doors to designing facile structural transformations in intercalation materials. For example, Schofield et al. \cite{parker2022} use a novel topochemical synthesis route to introduce dopants (e.g., Molybdenum) at selective sites of the cathode V$_2$O$_5$ compound. By doing so, they not only tailor its lattice parameters with a precision of $\sim0.01$\AA, but also can accurately design a facile structural transformation mitigating its mechanical degradation with repeated use \cite{markus2014computational,luo2022cation, parker2022}. These techniques have led to the development of novel compounds with reduced structural distortions and are promising candidates to develop zero-strain electrodes \cite{oh2021recent}. 

These experimental efforts to crystallographically design novel compounds are often complemented by first principle calculations \cite{yang2021synchronous, kolli2021six, kaufman2019understanding}. The first principle calculations not only provide crucial insights into the structural transformation mechanisms and the chemical stability of doped compounds, but also provide a framework to computationally design novel compounds. For example, a first principles method was used to systematically study how transition metal substitutions in layered oxide compounds would enhance battery related properties (e.g., thermodynamic stability, ionic mobility). This computational approach of chemically designing intercalation compounds could be used as a precursor to guide the experimental synthesis of new compounds. Recent developments in high-throughput computing \cite{tabor2018accelerating, jain2013commentary, saal2013materials} has led to an enormous database of intercalation materials, for e.g., The Materials Project, Inorganic Crystal Structure Database, which further guide experimentalists to synthesize new compounds \cite{jain2013commentary, zagorac2019recent}. These advances in theoretical and experimental methods provide invaluable tools to design structural transformations in intercalation materials and accelerate the discovery of novel compounds.

\begin{figure}
    \centering
    \includegraphics[width=0.9\textwidth]{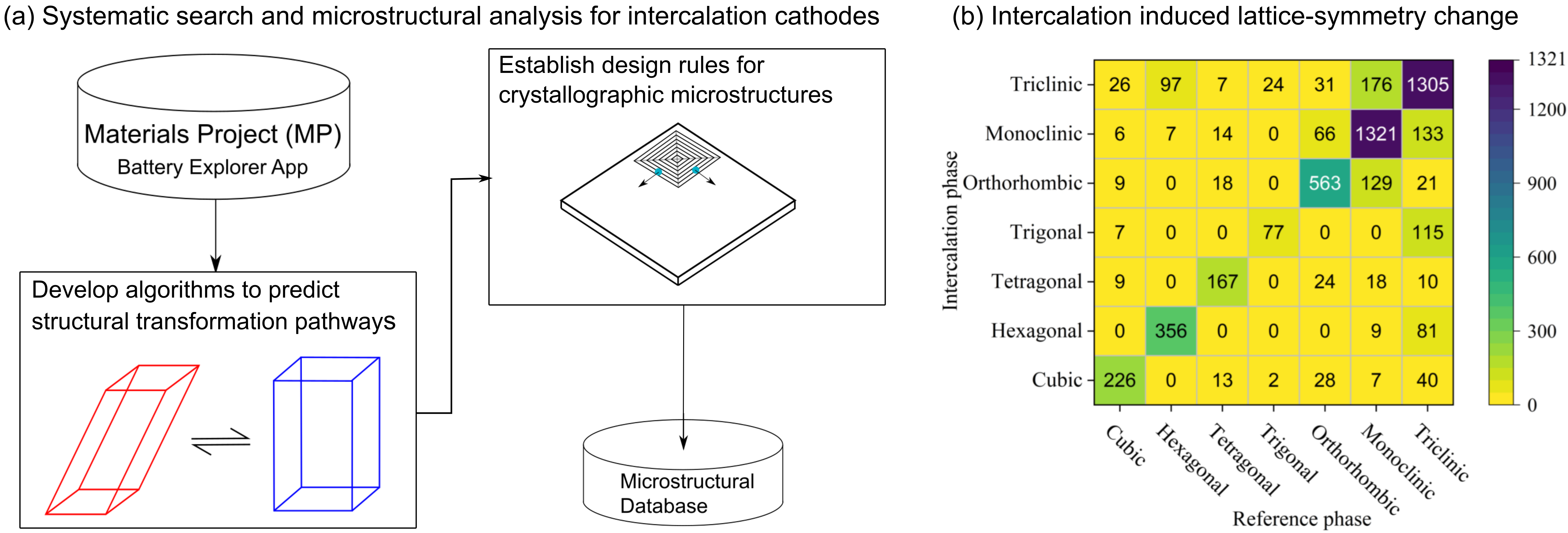}
    \caption{(a) We analyze structural transformations in over 5,000 intercalation compounds documented on the Materials Project database, and identify candidate compounds that satisfy crystallographic design rules to form self-accommodating or highly-reversible microstructures. (b) Our analysis shows that only $\sim 10\%$ of known intercalation compounds undergo lattice-symmetry lowering transformations, and thus can form twin boundaries and other crystallographic microstructures. Reprinted from \cite{zhang2021systematic} with permission from the authors.}
    \label{fig: Systematic-search}
\end{figure}

In an ongoing line of research in my group, we develop algorithms to quantify structural transformation pathways in commonly used intercalation materials, and establish crystallographic design rules necessary to reduce coherency stresses and volume changes in these materials, see Fig.~\ref{fig: Systematic-search}(a-b). We apply our analytical framework to over 5,000 intercalation compounds documented on The Materials Project database to identify candidate compounds that satisfy the crystallographic design rules. These candidate compounds and the crystallographic design rules would guide material chemists to systematically design structural transformation pathways in intercalation compounds. Besides the crystallographic design rules, there is a need to establish chemical substitution charts that identify stable compositions of intercalation compounds on introducing dopants. These charts would assist experimentalists in choosing suitable dopants to alloy intercalation compounds. With the growing need for increased energy densities and the urgency to reuse and recycle materials, it is important to pursue these combined theoretical and experimental approaches to crystallographically design novel compounds.\footnote{More recently, researchers are developing novel multivalent intercalation compounds that promise increased energy densities (exchange more than one ion) when compared to the conventional Li-ion batteries. Inserting these multivalent ions into host electrodes, however, are often accompanied by structural degradation because of the typically large ionic radius of the intercalating species. The widespread nature of these chemo-mechanical challenges further emphasize the need to crystallographically design materials with facile structural transformation pathways.}

\section*{Microstructures}\label{section: Microstructures}

\begin{figure}
    \centering
    \includegraphics[width=0.7\textwidth]{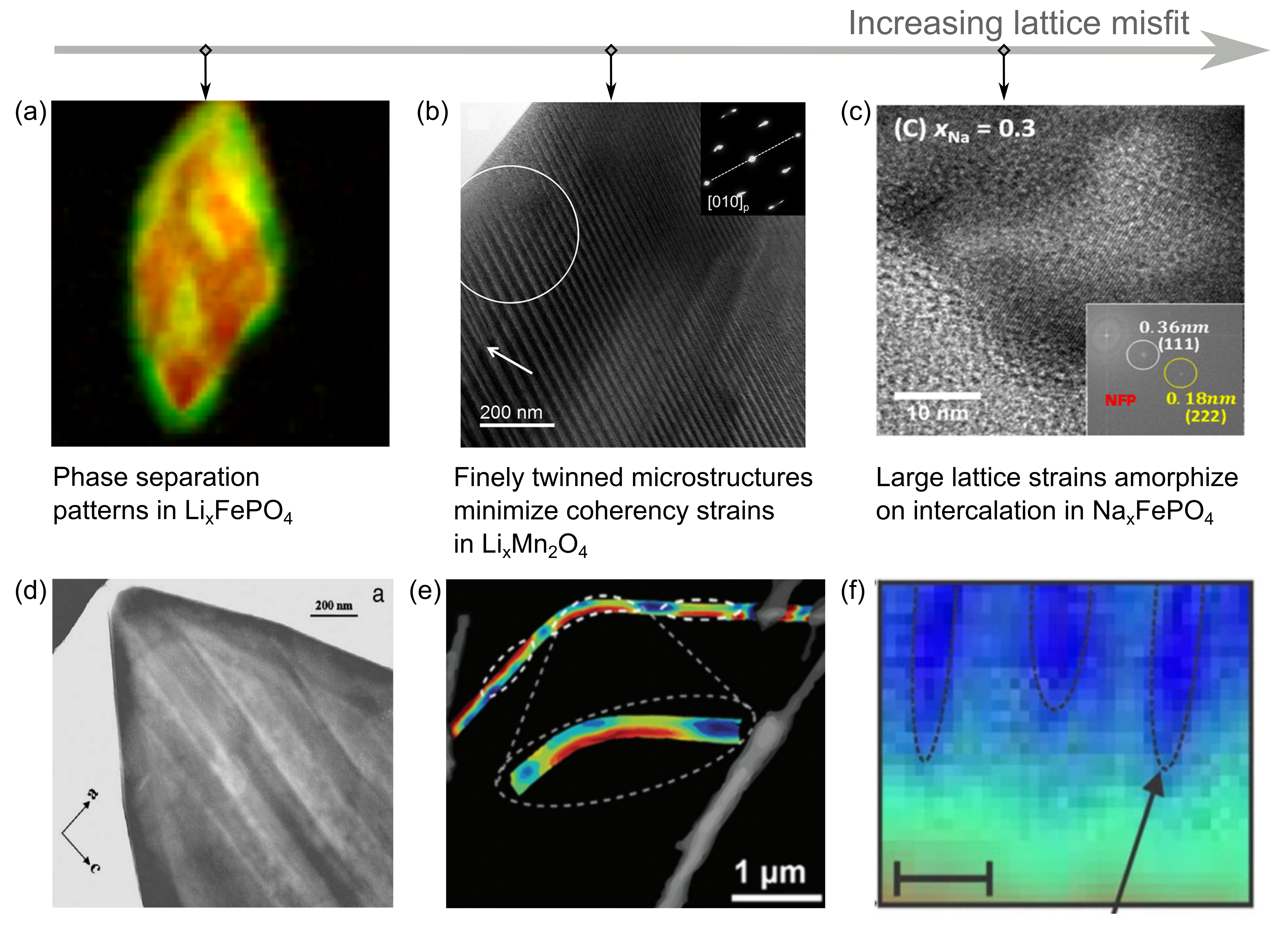}
    \caption{Phase separation microstructures form to reduce the chemo-mechanical energy of the system. Intrinsic lattice misft generates a wide range of microstructures: (a) For small lattice misfit spinodal decomposition patterns (such as in Li$_x$FePO$_4$) forms during intercalation. From \cite{lim2016origin}. Reprinted with permission from AAAS. (b) Other microstructures such as finely twinned martensite-like mixtures form in Li$_x$Mn$_2$O$_4$ to reduce the coherency stresses between lithiated and delithiated phases. Reprinted with permission from \cite{erichsen2020tracking} and ACS. (c) For extremely large lattice misfit ($>20\%$) a disordered phase such as in Na$_x$FePO$_4$ forms during intercalation. Reprinted with permission from \cite{xiang2017accommodating} and American Chemical Society. Other external constraints drive microstructural patterns: (d) Electrode particle size (and surface energy effects) generates a stripe-like microstructural pattern in Li$_{0.5}$FePO$_4$; Reprinted with permission from \cite{chen2006electron}. (e) Bending deformation stresses drives phase separation patterns in Li$_x$V$_2$O$_5$; Reprinted with permission from \cite{santos2020bending}. (f) Epitaxial strains generate a wave-like pattern of the intercalation front in Li$_x$FePO$_4$. Reprinted with permission from \cite{ohmer2015phase} and Springer Nature.}
    \label{fig: Microstructures-1}
\end{figure}

The abrupt structural transformation of individual lattices in intercalation electrodes generates significant misfit between neighboring lattices. This lattice misfit manifests at the continuum scale as coherency strains, which in turn nucleate defects and anisotropically expand electrodes, leading to their irreversible damage. Intercalation materials minimize these coherency stresses by forming microstructural patterns. These microstructures can range from a simple two-phase pattern to a complex finely-twinned pattern, see Fig.~\ref{fig: Microstructures-1}(a-c) \cite{li2014current, erichsen2020tracking}. Besides structural transformations, applied boundary conditions on the electrode and engineering design constraints drive the formation of rich and varied microstructural patterns in intercalation systems, see Fig.~\ref{fig: Microstructures-1}(d-f) \cite{santos2020bending, zhang2020stress, mistry2020stochasticity, cho2015influence, chen2015probing}. How these microstructures nucleate and grow affects rate capabilities and lifespans of intercalation materials \cite{lim2016origin}. In this section, I review the origin of these phase-separating microstructures in intercalation electrodes and highlight potential opportunities to crystallographically design these microstructures to enhance material longevity and performance.

\subsection*{Origin of phase separation microstructures}
Phase separating microstructures are characterized by a two-phase coexistence---within a single electrode particle---comprising of a Li-rich region (lithiated phase) and a Li-poor region (delithiated phase). These regions are separated by a phase boundary that moves back and forth through the electrode during charge / discharge cycles. Fig.~\ref{fig: Microstructures-1}(a-f) shows the types of microstructural patterns that form in intercalation electrodes. In this subsection, I review three key factors that drive the formation of these phase separation microstructures.

First, are the coherency strains that arise from misfit between neighboring lattices. These coherency strains increase the elastic energy of the system and phase separation microstructures form at the continuum scale to minimize these coherency strains, see Fig.~\ref{fig: Microstructures-1}(a-c). In intercalation electrodes with facile transformations (i.e., minimum crystallographic changes on Li intercalation), such as Li$_x$FePO$_4$, the phase boundary has negligible coherency strains and consequently forms phase separating microstructures that resemble the spinodal decomposition patterns, see Fig.~\ref{fig: Microstructures-1}(a). However, in intercalation electrodes with appreciable coherency strains, such as Li$_x$Mn$_2$O$_4$, the phase boundaries are elastically stressed. To reduce this elastic stress, the lithiated phase (Li$_2$Mn$_2$O$_4$) forms a mixture of finely twinned domains that has an average lattice deformation compatible with the delithiated phase (LiMn$_2$O$_4$), see Fig.~\ref{fig: Microstructures-1}(b) \cite{erichsen2020tracking}. These microstructures reduce the coherency stresses at the phase boundary, but at the same time act as a barrier for reversible transformation, leading to irreversible capacity loss of the material \cite{thackeray2018quest, rossouw1990structural}.\footnote{These finely twinned microstructures in spinel compounds bear a striking resemblance to the characteristic martensitic microstructures formed by energy minimization in shape memory alloys \cite{chu1995analysis}. In our recent work, we draw on insights from phase transformation materials such as shape-memory alloys to crystallographically design the phase separation microstructures in intercalation electrodes \cite{zhang2021systematic}.} Finally, in intercalation electrodes with very large lattice misfit (or coherency strains) phase separation is suppressed resulting in a solid-solution type of microstructure \cite{bai2011suppression, cogswell2012coherency}. In extreme cases, such as NaFePO$_4$ the large lattice misfit leads to amorphization of the intercalation compound \cite{xiang2017accommodating}, see Fig.~\ref{fig: Microstructures-1}(c). 

Second, the polycrystalline texture of intercalation electrodes affects the evolution of phase separating microstructures. For example, certain crystallographic orientations facilitate faster diffusion of the intercalating species---this anisotropic diffusion leads to heterogeneous lithiation and charge distribution in the host electrode \cite{ryu2018capacity}. Additionally, the random orientation of grains in a polycrystalline electrode creates tortuous diffusion pathways that impede charge/discharge rates. In another example, the heterogeneous lithiation induces inhomogeneous volume changes of the polycrystalline electrode particle. This inhomogeneity generates intergranular stresses that nucleates microcracks and contributes to the irreversible degradation of the electrode \cite{xu2018chemomechanical, ebner2013visualization, woodford2010electrochemical}. Researchers have developed theoretical and experimental tools to crystallographically enhance the performance of polycrystalline electrodes. Recently, Xu et al. \cite{xu2020charge}, demonstrated that architecting the crystallographic orientations of grains in a radial pattern contributes to smaller charge heterogeneity and reduced intergranular stresses in a polycrystalline electrode, see Fig.~\ref{fig: DefectEngineering}(a). Similarly researchers have identified grain morphologies and transformation pathways that mitigate structural degradation of electrode particles.

Third, the applied mechanical constraints, such as electrode particle geometry, substrate straining, and curvature induced stresses, drive the formation of microstructural patterns. For example, the platelet-shaped electrodes induce phase separation in Li$_{0.5}$FePO$_4$ compounds to form stripe-like patterns, see Fig.~\ref{fig: Microstructures-1}(d). In another example, the rod-shaped electrodes (that have enhanced ion-exchange because of their inherent geometry) are often bent in their secondary electrode configuration \cite{andrews2020curvature, santos2020bending}. This bending mode of deformation stresses the electrode particle and affects the phase separation microstructural pattern. Fig.~\ref{fig: Microstructures-1}(e) shows how the Li-rich/Li-poor phases in a rod-like V$_2$O$_5$ electrode separate into a periodic pattern to reduce the imposed bending constraint. Similarly, thin film electrode geometries are often constrained by a rigid substrate, which destablizes the intercalation-front \cite{zhang2020stress}. For example, tensile stresses in LiFePO$_4$ generates a wave-like pattern of the intercalation-front leading to inhomogeneous lithiation of the electrode, see Fig.~\ref{fig: Microstructures-1}(f).\footnote{In another line of research, mechanical constraints such as substrate strains have been used to delay the onset of large structural transformations in thin film electrodes \cite{zhang2021film}.} Overall, the microstructural patterns formed in intercalation electrodes are affected by physical phenomena spanning multiple length scales that range from structural transformations of individual lattices at the atomistic scale to concentration inhomogeneities at the continuum scale. 

\subsection*{Designing microstructural patterns}

How phase transformation microstructures nucleate and grow in intercalation materials greatly affects their properties. For example, the wave-like microstructural patterns that form under tensile stresses lead to a non-uniform lithiation of the electrode; this non-uniform lithiation can be detrimental contributing to current hotspots and mechanical degradation. Similarly, the phase separating microstructures in LiFePO$_4$ generate significant coherency stresses during charging/discharging; these coherency stresses can nucleate microcracks, and with repeated intercalation, contribute to material failure. It is important to engineer microstructural patterns to improve material performance and lifespans.

To date researchers have developed several techniques to design and control microstructural evolution in intercalation materials. For instance, Lim et al. \cite{lim2016origin} control the electrochemical operating conditions, such as diffusion kinetics, to suppress phase separation microstructures and to instead stabilize a solid solution pattern. This approach minimizes coherency stresses and retains the specific capacity of LiFePO$_4$. In another study, Santos et al. \cite{santos2020bending} impose mechanical constraints, such as engineering epitaxial strains and imposing bending stresses, to modulate the nucleation and growth of phase separation microstructures. This approach demonstrates how mechanical constraints are an additional design parameter that can be engineered to control the nucleation and growth of microstructural patterns. In addition to designing phase transformation microstructures in intercalation cathodes, researchers engineer electrode particle geometries (e.g., by faceting electrodes along specific crystallographic planes) and the underlying crystallographic texture (e.g., by designing grain orientations) to facilitate rapid charge/discharge capabilities. These approaches to designing microstructural patterns in intercalation electrodes has improved its electrochemical performance and extended its lifespans.

In addition to these ongoing efforts, I propose a new line of research in which we \textit{crystallographically design} microstructures in intercalation materials to achieve a dramatic improvement of material properties. Although phase transformation microstructures evolve at the continuum length scale, their origin is closely related to how individual lattices deform at the atomic scale. This exact correspondence between continuum microstructures and atomic lattice deformations is formalized via a Cauchy-Born rule that forms the basis for this proposed line of research \cite{ericksen2008cauchy}.\footnote{The Cauchy-Born rule is used together with the energy minimization theory to explain the origin of microstructures and more recently to design microstructural patterns that are stress-free and are highly reversible.} This rule has been successfully applied to crystallographically design microstructures in other phase transformation materials---such as shape memory alloys, ferromagnets, and ferroelectrics---which bear striking similarities to intercalation compounds (e.g., finely twinned microstructures).

For example, Chluba and co-workers, systematically dope a Ti$_{51}$Ni$_{36}$Cu$_{13}$ shape memory alloy to satisfy a specific combination of lattice parameters for which a stress-free phase boundary forms \cite{chluba2015ultralow}. This alloy typically develops fatigue after 200 stress-strain cycles, however by crystallographically designing its microstructure (via doping), researchers discovered a novel alloy composition Ti$_{54}$Ni$_{34}$Cu$_{12}$ that shows negligible fatigue despite cycling it for ten million times. Similarly crystallographic designing of microstructures has been used to engineer self-accommodating patterns (which accommodate with the macroscopic material shape and have zero volume changes) and twin interfaces, and this strategy has been applied to materials beyond shape memory alloys \cite{wegner2020correlation, pang2019reduced, liang2020tuning}. These results highlight potential opportunities for crystallographic designing of  microstructures in intercalation materials.\footnote{Here, crystallographic designing refers to systematic doping of intercalation compounds to tailor lattice geometries and stabilize shape-memory-like microstructures.}

These advances in the crystallographic designing of materials would need to complemented by developments in theoretical and experimental frameworks. For example, it is important to understand of how the crystallographic texture of the intercalation material interacts with the diffusing species during intercalation. At present, material models typically describe microstructures in intercalation materials as a function of a continuum field variable (e.g., Li-composition) \cite{wang2020application}. These models provide important insights into the structure of a phase boundary \cite{tang2011anisotropic}, interfacial dynamics \cite{singh2008intercalation}, and stress evolution \cite{zhang2020mechanically} in intercalation materials. These models, however, do not account for the underlying heterogeneous lattice deformations in a polycrystalline material. To address this gap, we will need a rigorous mathematical framework that couples the interaction between the crystallographic texture of the intercalation material and the Li-composition diffusion. Previously developed phase-field frameworks for chemo-mechanical materials, such as combining the Cahn-Hilliard model with a phase-field-crystal model, would be an initial step in this direction \cite{balakrishna2018combining, balakrishna2019phase}. These mathematical models would reveal crucial insights into the origin of crystallographic microstructures, and assist experimentalists to probe microstructural evolution in intercalation compounds.

\section*{Defects}\label{section: Defects}

\begin{figure}
    \centering
    \includegraphics[width=0.5\textwidth]{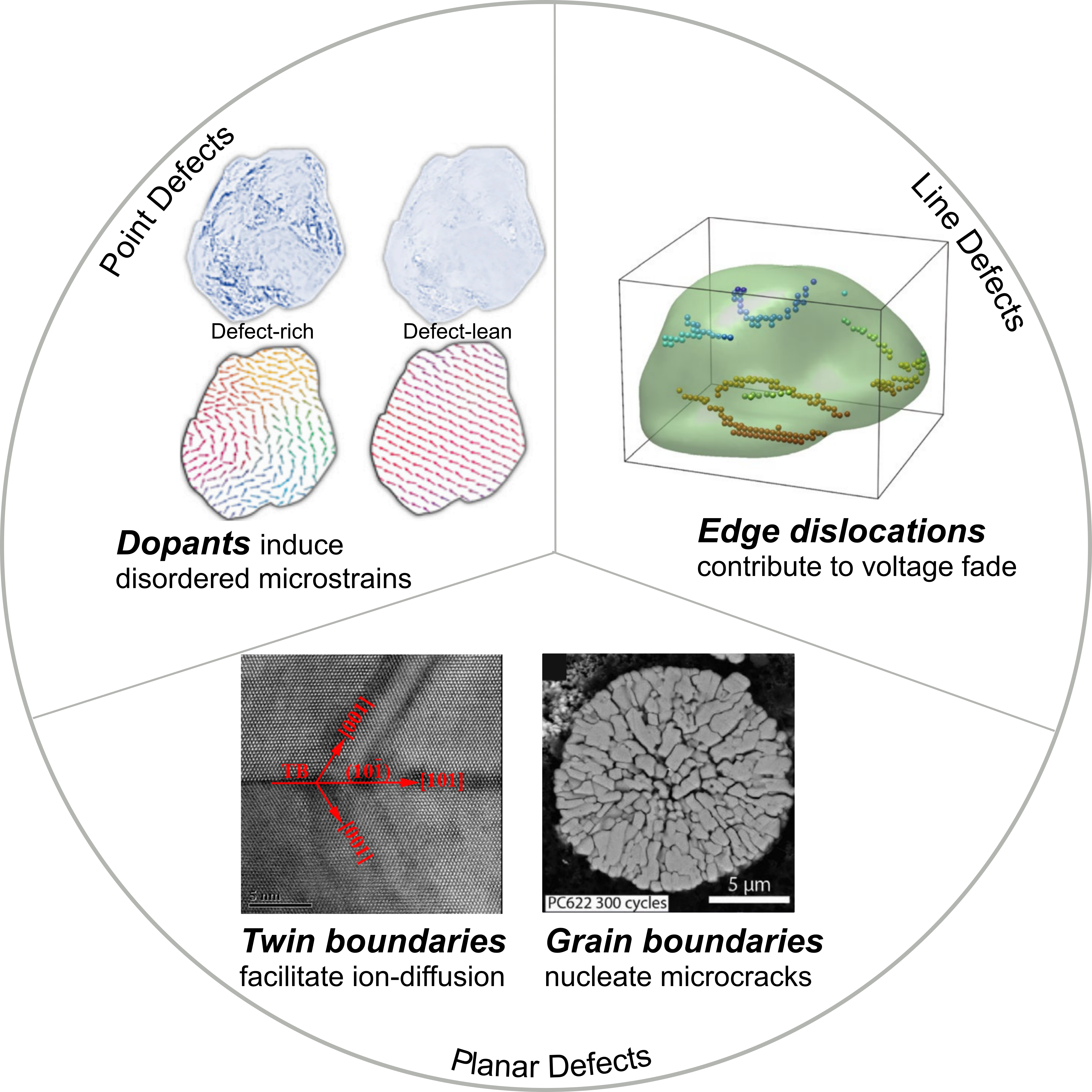}
    \caption{Commonly observed crystal defects in intercalation compounds can be categorized into (a) point defects (e.g., dopants, vacancies), (b) line defects (e.g., edge-dislocations), and (c) planar defects (e.g., twin boundaries, grain boundaries). These crystal defects can form either during manufacturing or during electrochemical cycling and can adversely affect the performance of the battery. Reprinted by permissions from \cite{hong2020hierarchical, singer2018nucleation, nie2015twin, qian2020single}.}
    \label{fig: Defects}
\end{figure}

Crystallographic defects are interruptions to the regular periodic crystal lattice and are widely observed in intercalation electrodes. These defects form during synthesis, processing, or electrochemical cycling of intercalation materials, and have been shown to have significant impact on its mechanical and transport properties. On the one hand, dislocation defects that nucleate during electrochemical cycling perturb the sequential arrangement of crystallographic planes and contribute to the electrode's irreversible voltage fade \cite{singer2018nucleation}. On the other hand, crystallographic defects such as clusters of vacancies can enhance electrode conductivity by mediating Li diffusion in intercalation electrodes \cite{van2013understanding}. In this section, we consider three families of crystallographic defects---namely point, line, and planar defects---and review how they affect an intercalation compound's chemo-mechanical properties, see Fig.~\ref{fig: Defects}. We conclude this section by identifying challenges and opportunities in engineering defects in intercalation materials and improving material properties.

\subsection*{Types of crystallographic defects}
Point defects are zero-dimensional atomic imperfections that occur only at a single lattice point and common examples include vacancies, dopants, and antisite defects. These defects, although considered as deleterious features, can have conducive effects on ionic mobility, lattice structural transformations, and phase transformation kinetics. For example, the antisite defects in LiFePO$_4$ block Li movement along the [010] migration channel, however simultaneously increase Li mobility along other crystallographic directions, contributing to an order-of-magnitude improvement in phase transformation kinetics \cite{hong2019mechanism}. Another example, is the trace doping of a cation (Ti) into an intercalation compound (e.g., LiCoO$_2$). These dopants segregate at particle surfaces and/or at grain boundaries and induce microstrains in the electrode, see Fig.~\ref{fig: Defects}. These internal strains suppress phase separation and facilitate longer lifespans of layered compounds \cite{hong2020hierarchical}.

Line defects are one-dimensional imperfections that correspond to a misaligned arrangement of atoms. A widely observed line defect is an edge-dislocation defect that corresponds to an abruptly terminated plane of atoms in the material bulk. These dislocation defects nucleate during electrochemical cycling and are found to be mobile \cite{singer2018nucleation, ulvestad2015topological}. For example, in layered oxide electrodes the nucleation and migration of dislocations disrupt the stacking sequence of layers in the host electrode. These structural changes are shown to block Li migration pathways and contribute to the voltage fade of the cathode particle \cite{singer2018nucleation}. In other electrodes, such as the spinel compounds, dislocations nucleate in the vicinity of a phase boundary to relieve coherency stresses. Although dislocation defects mitigate coherency stresses via plastic deformation, excessive accumulation of dislocations contributes to the irreversible transformation of the electrode \cite{erichsen2020tracking, yan2017intragranular}.

Planar defects are two-dimensional and typically correspond to interfacial regions separating domains with differing crystallographic orientation. Common examples of planar defects include grain boundaries, twin boundaries, and stacking faults. These defects can have both positive and negative impacts on material properties. One the one hand, planar defects such as twin boundaries are shown to have lower barrier for ionic diffusion and thus facilitate faster lithium transport. Engineering these planar defects in intercalation cathodes, such as spinel compounds, has lead to its fast charging properties \cite{nie2015twin, wang2021twin}. On the other hand, planar defects such as grain boundaries are shown to be sites of microcracking and fracture in polycrystalline electrodes. For example, the anisotropic volume changes of individual grains during intercalation generates interfacial stresses at the grain boundary, which at critical values leads to inter granular fracture \cite{besli2018mesoscale}. These fractured interfaces lead to the loss of ionic contact and contribute to the local impedance in cathodes. 

Crystal defects---point, line, or planar---commonly occur in intercalation materials and/or nucleate on repeated intercalation. These defects interact with Li-diffusion front during charging/discharging processes and can adversely impact a material's mechanical and transport properties. 

\subsection*{Designing crystallographic defects}

In this subsection, I argue that although crystal defects are deleterious features of intercalation materials impeding electrochemical performance, select crystal defects can be crystallographically engineered to enhance the material's mechanical and transport properties. I present case studies in which researchers have engineered defects---anti-site defects (point defect) and grain boundaries (planar defects)---to improve intercalation electrode performance. I will conclude this subsection by identifying open questions and opportunities to systematically design crystal defects in intercalation materials.

Anti-site defects are point defects that are generated when a cation (that is typically less mobile) occupies the site of an intercalating ion. These anti-site defects block the ion-diffusion pathways in intercalation materials and can negatively impact its transport properties \cite{kim2015effect}. One approach is to remove these anti-site defects by structural annealing of electrodes and to open the ion-diffusion pathways \cite{rasool2019mechanochemically}. By contrast, another approach is to engineer anti-site defects to enhance ion movement along other diffusion pathways and therefore contribute to a collective improvement of material's rate capability. Tang and co-workers, use a phase-field framework to show how intercalation materials with anisotropic diffusion (i.e., ion migrates predominantly along 1D channels) can be engineered with anti-site defects to enhance ion diffusivity along two-dimensions \cite{hong2019mechanism}. This improvement accelerates phase transformation rates by an order of magnitude and facilitates uniform surface-reactions in electrode particles. 

\begin{figure}
    \centering
    \includegraphics[width=0.3\textwidth]{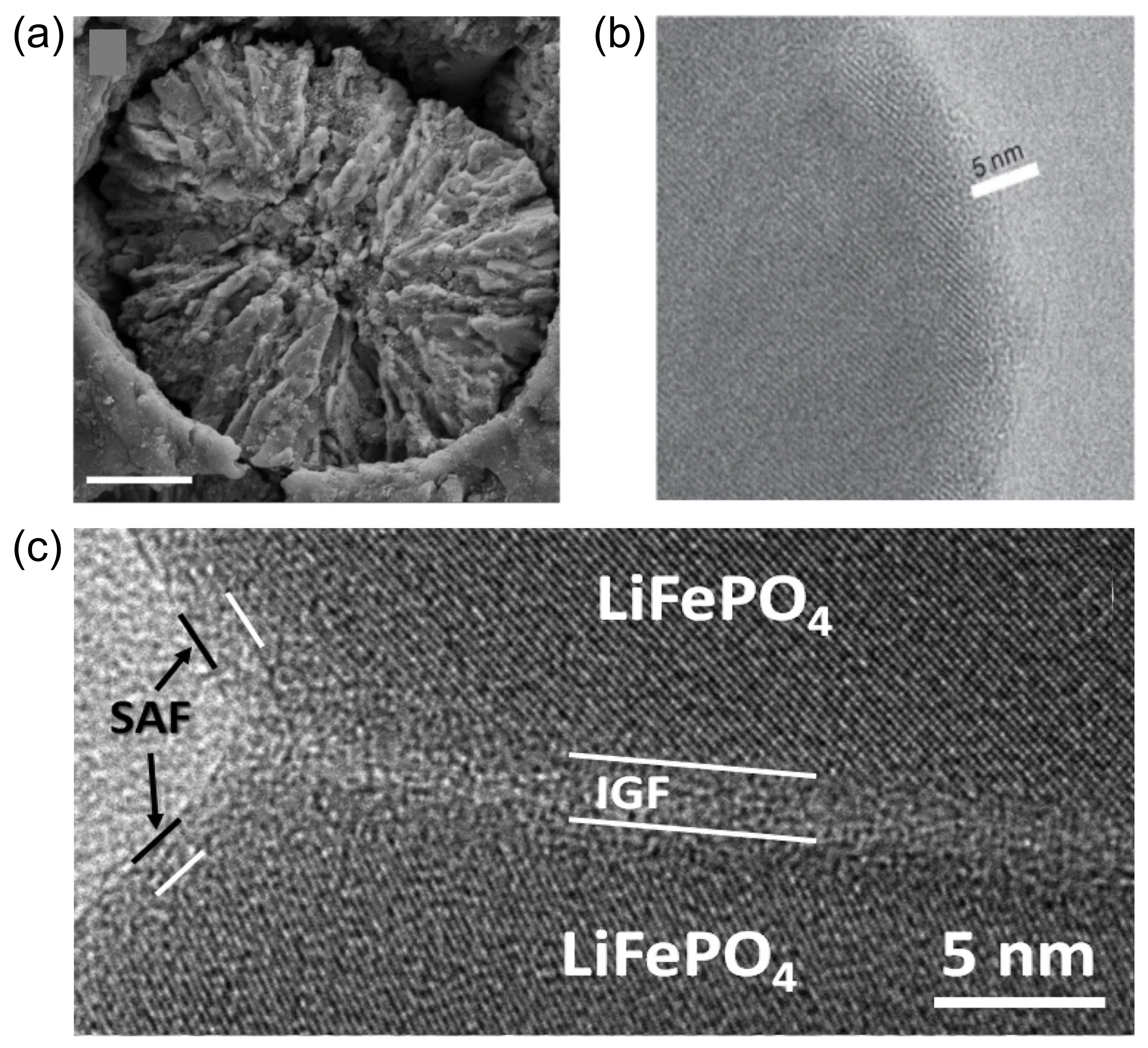}
    \caption{Select crystal defects can be engineered into intercalation compounds to improve its properties. (a) By designing a polycrystalline Ni-Mn-Co intercalation cathode with radially oriented grains Xu et al. \cite{xu2020charge} create less tortuous ion-diffusion pathways. Reprinted by permission from \cite{xu2020charge}. (b) An amorphous surfacial layer on LiFePO$_4$ compound facilitates faster adsorption and ion-exchange rate between electrode and electrolyte \cite{kang2009battery}. Reprinted by permission from \cite{kang2009battery} and Springer Nature. (c) Kayyar et al., \cite{kayyar2009surface} stabilize amorphous films along grain boundaries which facilitate faster ion-diffusion in intercalation compounds. Reprinted by permission from \cite{kayyar2009surface} and AIP publishing. }
    \label{fig: DefectEngineering}
\end{figure}

Grain boundary defects are ubiquitous in intercalation materials and, as discussed above, serve as sites for microcracking and fracture. While these chemo-mechanical challenges can be overcome by synthesizing high-quality single crystal particles, these synthesis methods often require high-temperature and pristine environments. Alternatively, planar defects, such as grain boundaries and surfacial films, can be thermodynamically engineered to achieve a dramatic improvement in cohesive and conductive properties, see Fig.~\ref{fig: DefectEngineering}(b-c). For example, Kang and Ceder stabilize a fast ion-conducting surface phase on LiFePO$_4$ particles, which increases its power rate by two orders of magnitude \cite{kang2009battery}. The disordered nature of the coating material on LiFePO$_4$ facilitates increased Li+ adsorption from the electrolyte. Similar disordered films have been stabilized by Luo and co-workers both on electrode particle surfaces and at their grain boundaries to greatly enhance charging/discharging capabilities. These surface films and intergranular films are abrupt transformations of crystalline interfaces and can be systematically designed (e.g., controlled processing conditions, introducing dopants, architecting grain boundary morphologies) to optimize battery material properties \cite{kayyar2009surface}.\footnote{In another line of research grain boundary complexions are being applied to anode electrodes to suppress the formation of solid-electrolyte interphase \cite{yan2020perspective}.}

These case studies demonstrate how small changes to the structure and composition of electrodes can dramatically improve their properties, however, it is unclear on how to experimentally design and optimize these crystal defects. For example, anti-site defects accelerate phase transformation rates in electrode particles, however, we do not know what concentration of anti-site defects and/or electrode particle geometries that would enhance electrode properties. Similarly, grain boundary complexions dramatically alter a material's cohesive and transport properties, however, the precise thermodynamic-and-kinetic conditions and grain boundary morphologies for which these complexions can be stabilized are not well understood. These examples highlight a need to establish design charts, such as complexion phase diagrams and defect density charts, to systematically engineer crystal defects in intercalation materials \cite{tang2010electrochemically}. For example, Luo and co-workers use advanced experimental methods, such as X-ray photo-electron spectroscopy and High-Resolution Transmission Electron Microscopy (TEM), to characterize interfacial complexions \cite{luo2019let}. These experimental techniques provide insights into the mechanism of grain boundary transformation and the stability of complexions in electrochemical environments. Furthermore, advances in large-scale physics-based simulations and data-driven methods are promising tools to elucidate the delicate interplay between grain boundary morphologies, thermodynamic variables, and kinetic conditions \cite{yan2022thermodynamics}. These studies contribute to establishing complexion phase diagrams, which would serve as design charts and guide crystal chemists to systematically engineer interfacial defects in intercalation materials \cite{cantwell2020grain}.

\section*{Summary}\label{section: Summary and Outlook}
In summary, intercalation materials hold great promise for reversible energy storage, but are plagued by chemo-mechanical challenges that limit their performance and lifespans. In this review, I outline these challenges stemming from structural transformations of individual lattices, collective evolution of phase transformation microstructures, and nucleation and growth of crystal defects. To mitigate these challenges, I identify potential opportunities of how we can intervene and crystallographically design intercalation compounds to enhance their properties and lifespans. I hope that these opportunities would inspire new solutions to the crystallographic designing of intercalation materials.
\newpage
\bibliographystyle{unsrt}
\bibliography{reference}

\end{document}